\documentclass[format=sigconf,10pt,letterpaper]{acmart}
\renewcommand\footnotetextcopyrightpermission[1]{} %
\pdfoutput=1

\usepackage{times}  
\hypersetup{pdfstartview=FitH,pdfpagelayout=SinglePage}
\usepackage{booktabs}  %
\usepackage{pbox}
\usepackage{subfigure}
\usepackage{epsfig,endnotes}
\usepackage{epstopdf}
\usepackage{xcolor}
\usepackage{graphicx}
\usepackage{xspace}
\usepackage{enumitem}
\usepackage{url}
\usepackage{breakurl}
\usepackage{balance}
\usepackage{amssymb} %
\usepackage{pbox}
\newif\ifcomment
\commenttrue
\ifdefined\nocomment
    \commentfalse
\fi

\ifcomment
    \newcounter{MVNumberOfComments}
    \stepcounter{MVNumberOfComments}
    \newcommand{\mvnote}[1]{\textcolor{blue}{\small \bf [MV\#\arabic{MVNumberOfComments}\stepcounter{MVNumberOfComments}: #1]}}
    \newcounter{ANNumberOfComments}
    \stepcounter{ANNumberOfComments}
    \newcommand{\anote}[1]{\textcolor{orange}{\small \bf [Ant\#\arabic{ANNumberOfComments}\stepcounter{ANNumberOfComments}: #1]}}
     \newcommand{\pnote}[1]{\textcolor{purple}{\small \bf [Panos\#\arabic{ANNumberOfComments}\stepcounter{ANNumberOfComments}: #1]}}

     \newcounter{IQNumberOfComments}
     \stepcounter{IQNumberOfComments}
     \newcommand{\iqnote}[1]{\textcolor{pink}{\small \bf [Inigo\#\arabic{IQNumberOfComments}\stepcounter{IQNumberOfComments}: #1]}}

    \newcommand{\NOTE}[1]
    {
      {\footnotesize\it
        \begin{center}
          \begin{tabular}{|c|}
           \hline
            \parbox{0.85\columnwidth}{
              \medskip
              #1
              \medskip} \\
            \hline
          \end{tabular}
        \end{center}
        }
    }
\else
    \newcommand\mvnote[1]{}
    \newcommand\anote[1]{}
    \newcommand\pnote[1]{}
    \newcommand\iqnote[1]{}    
    \newcommand\NOTE[1]{}
\fi

\newcommand{\eg}{{e.g.,}\xspace}

\newcommand{\ie}{{\it i.e.,}\xspace}
\newcommand{\folder}{./fig}

\newcommand\co[1]{}

\newcommand{\tool}{{$VPN^{0}$}\xspace}
\newcommand{\dvpn}{{dVPN}\xspace}

\newcommand{\pk}{\textsf{pk}}
\newcommand{\sk}{\textsf{sk}}
\newcommand{\sign}{\textsf{Sign}}
\newcommand{\messsigned}{m}
\newcommand{\verify}{\textsf{Verify}}
\newcommand{\signature}{s}
\newcommand{\enc}{\textsf{Enc}}
\newcommand{\dec}{\textsf{Dec}}
\newcommand{\ciphertext}{C}
\newcommand{\sni}{SNI}
\newcommand{\encryptedsni}{\ciphertext_{\sni}}
\newcommand{\SPK}{SPK}
\newcommand{\sproof}{\Pi}

\newcommand{\pkdomain}{\pk_{\domain}}

\newcommand{\pklastdht}{\pk_{\lastdht}}
\newcommand{\sklastdht}{\sk_{\lastdht}}

\newcommand{\encryptedkey}{\ciphertext_{\pkdomain}}
\newcommand{\signaturelastdht}{\signature_{\lastdht}}

\newcommand{\user}{S}
\newcommand{\relayer}{A}
\newcommand{\domain}{D}
\newcommand{\lastdht}{R}

\begin{document}

\title{$VPN^{0}$: A Privacy-Preserving Decentralized Virtual Private Network}
\author{Matteo~Varvello$\dag$, Inigo Querejeta Azurmendi$\dag\bullet$, Antonio Nappa$\dag$, Panagiotis Papadopoulos$\dag$, Goncalo Pestana$\dag$, Ben~Livshits$\dag\diamond$}
\affiliation{%
  \institution{$\dag$~Brave~Software, $\diamond$~Imperial~College~London, $\bullet$~UC3M}
}

\begin{abstract}
Distributed Virtual Private Networks (dVPNs) are new VPN solutions aiming to solve the trust-privacy concern of a VPN's central authority by leveraging a distributed architecture. In this paper, we first review the existing dVPN ecosystem and debate on its privacy requirements. Then, we present \emph{\tool}, a dVPN with strong privacy guarantees and minimal performance impact on its users. \tool guarantees that a dVPN node only carries traffic it has ``whitelisted'', without revealing its whitelist or knowing the traffic it tunnels. This is achieved via three main innovations. First, an attestation mechanism which leverages TLS to certify a user visit to a specific domain. Second, a zero knowledge proof to certify that some incoming traffic is \emph{authorized}, \eg falls in a node's whitelist, without disclosing the target domain. Third, a dynamic \emph{chain} of VPN tunnels to both increase privacy and guarantee service continuation while traffic certification is in place. The paper demonstrates \tool functioning when integrated with several production systems, namely BitTorrent DHT and ProtonVPN. 
\end{abstract}

\maketitle

\section{Introduction}
\label{sec:intro}
A Virtual Private Network (VPN) is a connection method used to add privacy to
private and public networks, like WiFi Hotspots or the Internet. Traffic
between the user (\emph{VPN client}) and a \emph{VPN node} is encrypted so that
network elements along the path have no access to this traffic. User's traffic is forwarded with the IP address of the VPN node, a feature many VPN providers offer as a remedy to geo-blocking. %

Users have to \emph{implicitly trust} VPN providers not to interfere with or log any of their personal traffic. It is to be noted that VPN providers are commercial entities
that might offer their services relying on other commercial entities (\eg they
could use multiple cloud services to obtain a worldwide footprint
\cite{ownvpn}). It follows that even trusted and respectable
vendors might unknowingly incur in issues with a specific provider ranging from
surveillance, misconfiguration, and even hacking. Either of these issues can
compromise users' privacy. In~\cite{khan2018empirical} the authors actively
investigate 62 commercial VPN providers and find unclear policies for non
logging, some evidence with tampering of their customer traffic, and a mismatch
between advertised VPN node locations and actual network location. 

Driven by the above issues, decentralized Virtual Private Networks (dVPNs)
arose as a fairly new trend with millions of daily users (\eg Hola~\cite{hola}). In a dVPN, users
are both VPN client and relay node as in a Peer-to-Peer (P2P) network. 
In spite of the apparent advantages on their privacy, users of dVPNs may need
to tunnel through their devices traffic that can be considered harmful or
illegal.  Indeed, there have been incidents
reported~\cite{holaAttack,holaDDos}, where unaware dVPN users have been
(ab)used as exit nodes through which DDoS attacks were performed. Similarly, the users have no guarantee on whether a dVPN might inspect, log, and share any of their traffic. 

In this paper, we first investigate the dVPN ecosystem and derive a set of \emph{requirements} from a privacy/performance standpoint. Next, we propose \tool, to the best of our knowledge the first privacy-preserving dVPN with traffic accounting  and traffic blaming capabilities. \tool is founded on the idea that dVPN nodes should be able to decide which traffic they want to carry, \eg only news websites. At the same time, they should accept such safe traffic in \emph{zero knowledge}, \ie without being able to tell what this traffic contains. Ultimately, we aim to offer the above features with minimum impact on the user experience.

We first note that such strong privacy guarantees are only possible in conjunction with already private traffic, \ie TLS v1.3~\cite{tls13} and DNScrypt~\cite{dnscrypt}. \tool further leverages a Distributed Hash Table (DHT) to pair dVPN clients with nodes currently available to serve their traffic. This pairing is realized using privacy preserving announce and lookup primitives. Most notably, we introduce VPN \emph{chains} which help both in preserving a user privacy and allowing uninterrupted VPN service. Last but not least, we introduce a zero knowledge traffic attestation mechanism piggybacking on TLS. 

We have integrated \tool --- to the extent that is possible without third party cooperation --- with  BitTorrent DHT~\cite{varvello2011traffic} and ProtonVPN~\cite{proton}, a popular VPN provider. We demonstrate the feasibility of \tool while directing testing to a public domain supporting TLS v1.3. We also benchmark \tool performance with respect to DHT lookup, VPN tunnel setup, and zero knowledge proof calculation. We identify the current bottleneck in the proof calculation, which makes our main avenue of future work.

\section{Motivation}
\label{sec:motivation}
VPN services enable users to bypass geo-blocking and enhance their privacy against snooping ISPs and malicious access points. To avoid the necessity of blindly trusting a centralized service provider that could harm their privacy (\eg, by logging their network connections), 
dVPNs came to the rescue. DVPNs are P2P VPNs where users forward their traffic through other users and vice-versa. %
Unfortunately, this architecture allows malicious users to abuse the network and perform malicious transactions via unaware users positioned as exit nodes.

We assume malicious dVPN users, hidden behind benign users' IPs, abuse the dVPN  by (i) accessing illegal content (\eg child pornography, darknet markets~\cite{Dittus:2018:PCL:3178876.3186094}), or (ii) launching distributed attacks against selected targets~\cite{holaDDos}. In addition, we assume a snooping ISP who logs the network traffic uploaded by the user's device. These logs can be used later for purposes beyond the control of the user (sold to advertisers~\cite{ispAdvertisers} or handed over to agencies~\cite{ispact}) that may result in tarnishing the user's reputation or even falsely accusing them for illegal transactions.

\subsection{Requirement Analysis}
\label{sec:requirements}
In this paper, we envision a dVPN service that will fulfill the following requirements:

  \vspace{0.1in}
  \noindent\textbf{IP Blacklisting:} To be usable, a VPN (both centralized and distributed) needs to publish at least a portion of its vantage point list. It follows that it is relatively easy for a censorship entity or a geo-blocking content provider to access such list and simply blacklist all the vantage points of a VPN. For centralized VPNs, this is an issue they constantly face and they can hardly solve. For dVPNs, such blacklisting is harder due to the dynamic set of users/IPs involved. DVPN nodes are regular Internet users who frequently change network  locations  and  connect  from  behind  Network Address Translators (NATs). In this case, blocking a NATed VPN node implies blocking the whole subnet with a potentially massive service disruption. 

  \vspace{0.1in}
  \noindent\textbf{No-Logging:} Privacy is a main service that should be offered by a VPN. This implies that, at no time, a VPN node should be able to log user traffic. In~\cite{khan2018empirical}, authors investigate the usage policy offered by several commercial VPNs on their website. They find that when a privacy policy was available (75\% of the cases), very few VPN services explicitly claimed a \emph{no-logs} policy. This analysis suggests that VPN providers today should do a better job in terms of transparency of their actions. In dVPN, logging is sometimes required to offer, for example, protection against IP blacklisting: in VPN Gate~\cite{vpnGate},  each VPN node keeps connection logs (and shares them with a central repository) to inform other VPN servers of a potential censorship authority attempting to discover (and block) the current dVPN footprint. 
  
  \vspace{0.1in}
  \noindent\textbf{Traffic Accounting:} The founding idea of a dVPN is that users share their resources, so there must be a system to account for such traffic and grant tokens accordingly. Crypto dVPNs~\cite{sentinel,mysterium} tackle this issue by leveraging the blockchain to keep track of proof of traffic. This can be challenging depending on which network logging level is allowed/required. 
  
  \vspace{0.1in}
  \noindent\textbf{Traffic Blame:} From a networking perspective, VPN nodes are the entity originating the traffic they carry. This means that serious offenses (\eg child pornography, hate speech, drug smuggling), when investigated, will point the authorities to the entity running the VPN service. At this point, the above no-logs policy comes into play where the VPN might (or not) offer extra information about who was indeed originating such traffic. In a dVPN context, there is no legal entity the authority can reach to. Instead, they would reach a victim dVPN user whose network was used to carry such traffic. It is thus paramount that a dVPN implements a mechanism to avoid this kind of situation. At the same time, this should be achieved in a privacy preserving way, thus respecting the above no logging requirement.

    \vspace{0.1in}
  \noindent\textbf{High Quality of Experience (QoE)} Offering high QoE is a hard task for dVPNs. This is because of client churn and heterogeneous network conditions; this problem is not specific to dVPNs but an overall generic issue in distributed systems. A VPN footprint, \ie how many unique locations a VPN can offer, is another important QoE metric. VPN providers constantly battle to offer more vantage points, either by deploying new physical nodes or by  introducing ``virtual locations'' based on the information available from geo-IP databases about the physical locations of their vantage points. One shared limitation among centralized VPNs  is the lack of residential IP addresses, since they mostly rely on data-centers to deploy their nodes.  Contrary to that, by definition, dVPNs consist of a large network footprint of residential IP addresses.

    \vspace{0.1in}
  \noindent\textbf{Open Source:} A dVPN client/server code is a very critical piece of software since it can potentially gain access to very sensitive data. Despite popular VPN tunneling protocols (OpenVPN and PPTP) are inherently secure, it is important to note that misconfigurations and/or malicious code are still potential threats~\cite{ikram2016analysis}.

\section{VPN-Zero: System Overview}
\label{sec:system}
This section presents the design of \tool, a privacy preserving dVPN based on \emph{zero knowledge}. A zero knowledge proof is a cryptographic tool that allows a \emph{prover} to prove to a \emph{verifier} that a certain statement is true, without disclosing any information except the fact that the statement validates. In our case, a dVPN user wants to prove to a dVPN node that the traffic it is sending is contained within the node's \emph{whitelist}, \eg a set of domains the node is willing to carry traffic for. 

Several challenges are involved to realize the above statement in a decentralized system. First, how to distribute such whitelists in a privacy preserving manner. Second, how to build a zero knowledge proof around \emph{traffic}, and implicitly which traffic is suitable for such proof. Last but not least, how to perform the above operations without completely disrupting the user experience. 

In the remainder of this section, we first set the stage for the traffic type \tool can operate on to guarantee strong privacy requirements. Next, we introduce the cryptographic primitives at the foundation of \tool. We then describe its distributed architecture along with the protocol orchestrating \tool's operations. Finally, we finish the section with a description of how we construct the zero knowledge proof.

\subsection{Foundations}
\vspace{0.1in}
\noindent\textbf{Confidential Traffic:} \tool's zero knowledge goal translates into strict privacy requirements for the traffic being carried. To this end, \tool can only be coupled with TLS v1.3~\cite{tls13} and DNScrypt~\cite{dnscrypt}. This implies a \dvpn node can only observe a destination IP address, along with TLS ClientHello and ServerHello messages. These building blocks allow \tool's users to carry traffic without identifying the target domain, contrarily to TLS v1.2~\cite{tls12} where the server domain name is sent in the clear during the handshake.

Further, even TLS v1.3 does not imply mandatory encryption of the Server Name Indication (SNI) field in the  ClientHello. This would allow a dVPN node to learn which domains a user is visiting. To prevent such privacy leak,  the Encrypted Server Name Indication RFC draft~\cite{snidraft} proposes a novel method to encrypt such information with the public key of the receiving server.  In \cite{snidraft} details are given of how to encrypt the SNI, and how the receiver of the TLS handshake only proceeds if such an encryption is performed with the correct public key. %

With respect to the destination IP, while this also can be used to learn
important information about the traffic, we here assume that orthogonal
solutions, \eg onion routing as in ToR~\cite{tor}, can be adopted. Nonetheless
IP addresses often do not disclose information about the destination, because
the servers are hosted in within cloud providers or sit behind a CDN.

\vspace{0.1in}
\noindent\textbf{Cryptographic Primitives:} \tool requires each of the nodes to own a public-private key-pair $(\pk, \sk)$.
We denote the signature on message $\messsigned$
with private key $\sk$ by $\sign(\messsigned, \sk)$. $\verify(\signature, \pk)$ verifies a signature
$\signature$ with public key $\pk$. If the signature is valid, it outputs $\top$. Finally we denote the encryption of message $\messsigned$ and
decryption of ciphertext $\ciphertext$ with $\enc(\messsigned, \pk)$,
$\dec(\ciphertext, \sk)$ respectively. 

\newcommand{\elgamal}{\textsf{ElGamal}}
\newcommand{\elgamalpk}{\pk^{EG}}
\newcommand{\elgamalsk}{\sk^{EG}}

Moreover, \tool relies on ephemeral one time key-pairs which are used each time
a client lookup a dVPN node. We use ElGamal key-pair
\cite{elgamal}, which results in a small overhead for the user (one
exponentiation over a finite field for each new key calculation once public parameters are computed for the whole network). We use
ephemeral keys so that connection requests cannot be linked to a particular user.
We use $\elgamal$ to denote the operations performed with the ElGamal keypair
$(\elgamalpk, \elgamalsk)$.

\begin{figure}[t]
  \centering
  \psfig{figure=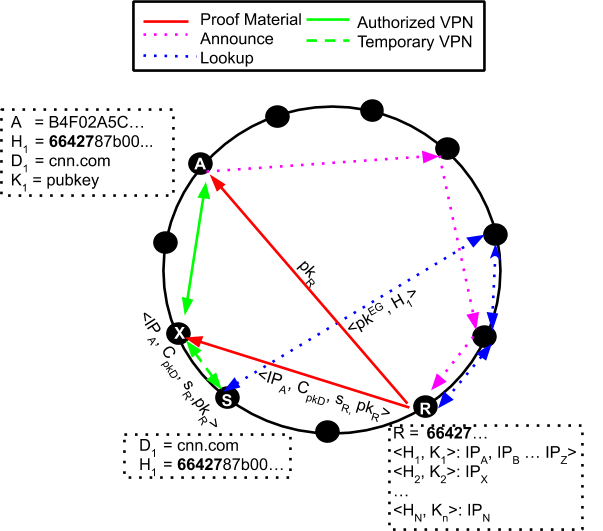, width=3.5in}
  \caption{Flow of a connection using \tool. First, nodes announce their whitelists. Then, when a user wants to connect to a domain, it first uses a temporary VPN, and then performs a lookup to find the node to create an authorized VPN with.}
  \label{fig:arch}
\end{figure}

\subsection{Network Architecture}
 \tool's architecture is built on top
of a Distributed Hash Table (DHT)~\cite{kaashoek2003koorde,zhao2004tapestry}.
The DHT is used to privately identify a set of \dvpn nodes willing to act as
VPN endpoints for some specific traffic. To this end, \dvpn nodes store in the
DHT each entry of their whitelists using the hash of the domain name as a key,
and as a value the domain's public key and their current public IP. 

Figure~\ref{fig:arch} shows an example (magenta dashed line) where node
$\relayer$ announces that it accepts traffic towards a destination
$\domain_{1}$. The hashing uniform distribution property enforces that a
whitelist is effectively scattered among multiple peers. This is important
since whitelists contain privacy sensitive information. Nodes frequently
re-publish their whitelist to account for fresh information and to update
reachability information, as commonly done in any DHT~\cite{kaashoek2003koorde,zhao2004tapestry}. A
classic time-to-live approach should be used to handle deletion of entries from
the DHT. 

When a user $\user$ wants to start a VPN session, it first opens a
\emph{temporary} VPN tunnel to some node $X$ (Figure~\ref{fig:arch}, green
dashed line). This can be, for instance, a recently used node. Next, $\user$
naturally originates some TLS (v1.3) traffic, \eg a visit to a secure domain
$\domain_{1}$. This temporarily \emph{unauthorized} traffic flows through $X$
for a duration $T$. Within $T$, $X$ will be able to locate 
a \dvpn node for which this traffic is authorized, if it exists.

Meanwhile, $\user$ performs a DHT lookup using $h_{1} = h(\domain_{1})$. We
assume an iterative lookup where each step routes the request to a node whose
DHT identifier is closer (bitwise) to $h_{1}$ (Figure~\ref{fig:arch}, blue
dashed line). Differently from a regular DHT lookup, $\user$ does not include
its IP address to the request but rather $X$'s IP so that the destination does
not learn which domain $\user$ wants to visit. Note that $X$ cannot perform the
DHT lookup directly since this would imply knowing $\domain_{1}$. Additionally,
$\user$ appends its ephemeral public key $\elgamalpk$ to the lookup. The later
converges to a node $\lastdht$ which returns to $X$ the IP address of a \dvpn
node, $\relayer$, accepting traffic to $\domain_{1}$ \footnote{Multiple selection
strategies can and should be investigated as a future work.}. Additionally, node
$\lastdht$ includes in the message to $X$ an encryption of the domain's public
key, $\encryptedkey = \elgamal.\enc(\pkdomain, \elgamalpk_{\user})$, together
with a signature of the latter, $\signaturelastdht = \sign(\encryptedkey,
\sklastdht)$ and its own public key $\pklastdht$. Finally $\lastdht$ connects with $\relayer$ to send
$\pk_{\lastdht}$, which allows $\relayer$ to verify that the accessed domain is
among its accepted domains.

Next, $X$ shares $\relayer$'s IP, together with $\encryptedkey,
\signaturelastdht$ and $\pklastdht$ with $\user$.  Meanwhile, $X$ opens a
temporary VPN tunnel to $\relayer$, the node selected to carry authorized
traffic to $\domain_{1}$. When the tunnel is ready, $X$ creates simple
\texttt{iptables} rules to connect the two VPN tunnels, effectively realizing a
VPN \emph{chain} ($S \rightarrow X  \rightarrow A$). 

The above VPN chain has an important effect on the existing TLS connection ($S
\rightarrow D$). From D's perspective, the endpoint was a socket at X
($IP_{X}:PORT$). Now, a new socket ($IP_{\relayer}:PORT$) is introduced. This
is similar to TCP hijacking~\cite{harris1999tcp}; however, in this case we
purposely do not attempt to hijack a TCP connection. The server will thus
realize, \eg wrong sequence number, and respond with a TCP RST. This will
travel back to S forcing a new TCP handshake and thus TLS handshake. 

The latter detail is very important since it implies that $\relayer$ will
observe a new TLS handshake. This allows $\user$ to compute a ZKP which
convinces $\relayer$ that the destination domain is the same which was
encrypted by $\lastdht$ in $\encryptedkey$. A potential avenue of attack here
is a collusion between $\user$ and $\lastdht$. This translates into a
\emph{sybil attack}, a popular attack on the DHT for which many countermeasures
exist~\cite{cholez2009evaluation}.  If the ZKP verification fails, the tunnel
is interrupted. Otherwise, this traffic is authorized without $\relayer$ ever
learning the domain's SNI or its public key. 

\subsection{Zero Knowledge Proof} 
During the forced re-negotiation of the TLS handshake, $\user$ sends again in
the ClientHello an encryption of the SNI under the domain's public key
$\encryptedsni = \enc(\sni, \pkdomain)$. This, together with the information
received from $X: \encryptedkey, \signaturelastdht$ and $\pk_{\lastdht}$ are the
key components of the ZKP used in \tool. 

In a nutshell, the user proves that the key used to encrypt $\encryptedsni$ is
the same than the one encrypted in $\encryptedkey$, and that
$\verify(\encryptedkey, \pklastdht)$ validates.  We adopt the Camenisch-Stadler
notation~\cite{Camenisch:1997:EGS:646762.706305} to denote such proofs and
write 

\begin{multline*} \sproof = \SPK\{ (\pkdomain, \sni, \elgamalsk_{\user}):\\
\encryptedsni = \enc(\sni, \pkdomain) \;\land\; \\
\elgamal.\dec(\encryptedkey, \elgamalsk_{\user}) = \pkdomain \; \land\; \\
\verify(\signaturelastdht, \pklastdht) = \top\} 
\end{multline*} 
to  denote the non-interactive signature proof of knowledge that the prover
  knows the public key $\pkdomain$ used to encrypt  $\sni$ in
  $\encryptedsni$, and encrypted in $\encryptedkey$, where the latter is
  signed by $\pklastdht$. The values between the parenthesis are kept private
   ($\pkdomain,
  \sni, \elgamalsk_{\user}$), while the other values
  used in $\sproof$ are public.
\vspace{0.1in}

Note that such a proof is not straightforward. We firstly prove that a
ciphertext, $\encryptedsni$, is the result of an encryption without disclosing
the public key nor the plaintext. This causes the highest overhead in our
construction. We use the construction presented in \cite{camenisch:safeprimes}
for this purpose. Then we need to link the public key encrypted in clause two,
with the one used in clause one. For this we use a proof that two commitments
hide the same secret \cite{interval:efficient}. Finally the third clause can be
openly computed by $\relayer$ given that it received the public key from
$\lastdht$.

Using this, $\user$ can convince $\relayer$ that the tunnel created is to a
domain that the latter considers valid, without disclosing which one.

\begin{figure}[t]
  \centering
  \psfig{figure=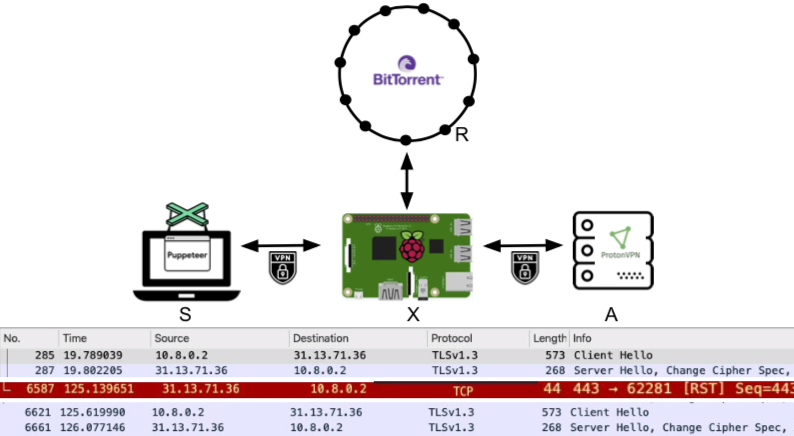, width=3.5in}
  \caption{High level overview of \tool's prototype.}  \vspace{-0.3cm}
  \label{fig:demo}
\end{figure}

\section{Preliminary Evaluation}
\label{sec:eval}
This section preliminary evaluates \tool. Apart from the zero knowledge calculation, \tool consists of well known components for which large scale production systems already exist. Instead of building a small scale testbed or some form of emulator/simulator, we have integrated \tool --- to the extent that is possible without third party cooperation --- with  Mainline~\cite{varvello2011traffic} (Bittorent DHT based on Kademlia with tens of millions of users), and ProtonVPN~\cite{proton}, a popular VPN provider. We further use OpenVPN~\cite{openvpn} to run the first VPN path on a Rasberry Pi we control. 

Figure~\ref{fig:demo} shows a graphical representation of our setup. Please note that we kept the node labeling from Figure~\ref{fig:arch} as a direct reference. The figure is further enhanced with wireshark data from the ongoing traffic captured at S. A headless browser~\cite{puppeteer} runs on a laptop (node A) and it is instrumented to visit a TLSv1.3-enabled website. For this test we used \texttt{facebook.com} (\texttt{31.13.71.36}), as shown in the bottom of the figure).  

The laptop connection is tunneled through a Raspberry Pi (node X) located in the same LAN as S; note that this is a worst case scenario for \tool since the extra network latency would further hide our traffic verification process. Whenever a new visit is started, node X performs a DHT query in Mainline. Note that in \tool design this operation is accomplished by A while spoofing X's IP address.  Since this would require a modification of \texttt{transmission}~\cite{transmission}, the BitTorrent Linux client we instrumented for our tests, we opted for a simplification which does not impact the performance evaluation. As input for the DHT lookups, we use the top 100 magnet links (DHT hashes) as indicated by ThePirateBay~\cite{thepiratebay}. As soon as the DHT response is received, our script opens a new tunnel to a random ProtonVPN server (node A) from the list provided with a basic subscription. Meanwhile, the ZKP is calculated at S to be then sent to A for tunnel validation. This latter step was not implemented since it would require collaboration with ProtonVPN.

\begin{figure}[t]
  \centering
  \psfig{figure=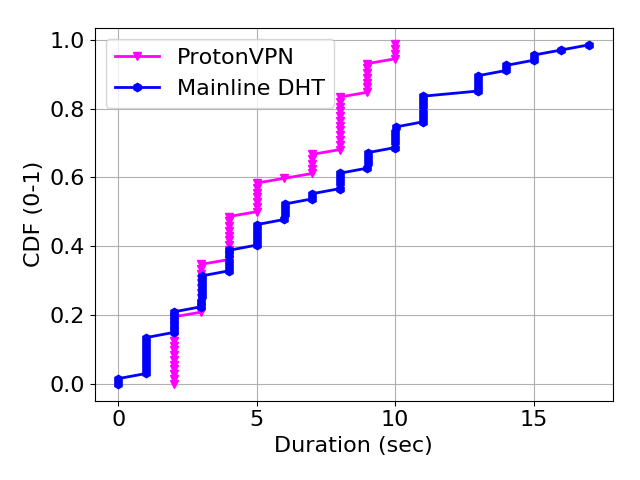, width=3in}\vspace{-0.3cm}
  \caption{CDF of lookup duration ; ProtonVPN and Mainline DHT.}  \vspace{-0.3cm}
  \label{fig:res-a}
\end{figure}

Figure~\ref{fig:res-a} shows the Cumulative Distribution Function (CDF) of DHT lookup duration in the Mainline DHT for its top 100 hashes. Overall, we observe a uniform distribution between 1 and 20 seconds. Contributions to this delay are: diverse network paths taken by DHT lookups, failures along the path, lookup replication factor, etc. It is worth noting that a similar result was measured for Mainline in~\cite{varvello2011traffic}, and~\cite{steiner2010evaluating} for KAD, the DHT used by eMule~\cite{emule}. Further, lookup optimizations are possible to speedup these operations, \eg Steiner et al.~\cite{steiner2010evaluating} show how KAD latency can be halved with no extra load on the DHT. 

The next operation post DHT lookup is the setup of the second leg of the VPN chain. Figure~\ref{fig:res-a} also shows the CDF of such VPN setup time computed for 72 VPN nodes --- 96 nodes were tested but 24 failed, \ie negotiation did not succeed within 30 seconds. This high failure rate is potentially due to a  ProtonVPN protection for too  frequent switches. The figure shows a median reconnect time of 4 seconds and worst case durations of up to 10 seconds. 

\begin{figure}[t]
  \centering
  \psfig{figure=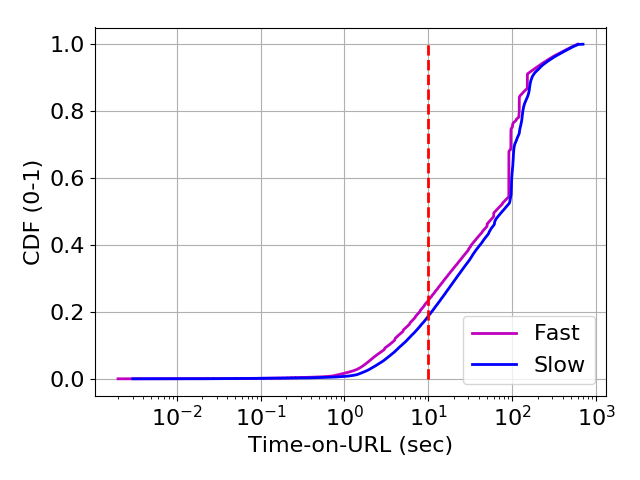, width=3in}\vspace{-0.3cm}
  \caption{CDF of time spent on URL ; 1M browser sessions.}  \vspace{-0.3cm}
  \label{fig:res-c}
\end{figure}

When putting all together, we estimate that \tool requires a median setup time of 10 seconds, and a worst time of 25-30 seconds. We benchmarked our ZKP calculation using a Python prototype implementation and measured an average of 10 seconds. This indicates that the ZKP calculation can potentially be \tool bottleneck, especially since both DHT lookup and VPN tunnel setup can be optimized, if needed. It is thus clear that our main future work consists in optimizing how the ZKP calculation should be carried to minimize its duration. Our current directions are both an improvement of the protocol, and a switch to better performing languages (Rust, C++). 

Note that the above latency is currently hidden to the user thanks to \tool design. However, the longer our procedure takes the higher the chance for the user to generate unauthorized traffic. To comment on the latter, we have analyzed one million browsing \emph{session} from Ciao~\cite{ciao_url}, a Chrome plugin which helps discovering and using free HTTP(S) proxies on the Internet. We define a session as the time spent at a specific URL, either manually entered in the browser or opened by clicking on a link. This consists of the page load time plus the actual time the user spends interacting with a page. Note that Ciao only collects page load time, time on site, and bytes transferred. Any other private information like IP address or URL requested are not collected. 

Figure~\ref{fig:res-c} shows the Cumulative Distribution Function (CDF) of the time spent on a URL for the above browsing sessions. We differentiate between `fast' and `slow' where `fast' does not account for the PLT which is slower than usual in our dataset since free proxies are used. We further enhance the figure with a vertical line showing \tool median verification time (10 seconds). The figure shows that 20\% of the sessions would change \emph{before} traffic verification. Note that this is a worst case analysis since our session definition potentially implies users remaining at a given domain, \eg by opening a new article on a news site.

\section{Related Work}
\label{sec:related}

\vspace{0.1in}
\noindent{\bf Zero Knowledge Proofs:} Guillou and Quisquater \cite{10.1007/3-540-45961-8_11} introduced a ZKP proof to verify RSA signatures, which could be easily extended to verify RSA encryptions under a certain key. However, for this construction, the knowledge of the public key (modulus and exponent) is necessary. To hide such information, we implemented Camenisch's ZKP of modular exponentiation presented in \cite{camenisch:safeprimes}. However, the range proof included in this paper is not up to date with current state of the art range proofs. Peng and Bao's \cite{Peng:2010:ERP:1906497.1907394} scheme improves on previous work, which is the construction we have used for \tool. 

\vspace{0.1in}
\noindent{\bf Distributed Virtual Private Networks:} The idea of a dVPN is far from novel, with many designs dating now more than 10 years back, \ie when P2P research was at its peak. To the best of our knowledge, ELA~\cite{1386101} was one of the first approach to decentralize a VPN. Several other variations then appear, such as SocialVPN~\cite{juste2010socialvpn} a system which drives the peer selection strategy based on social relationships among nodes, or N2N where users share a common encryption key obtained when they join. None of this approach provides strong privacy guarantees as \tool. However, we start noticing a trend towards protecting user identify and traffic, \eg only routing it through friends and providing strong online identities. 

In 2007, the first dVPN ``product'' was launched: Hola~\cite{hola}. Hola is a freemium web and mobile application which attracted, over the years, tens of millions of users/nodes. In Hola, users agree to either pay a premium per month or offer part of their upload bandwidth to other Hola users. Several concerns have been raised about the security of Hola~\cite{adiosHola} and there is a lot of criticism about the fact that  many free users are unaware that their bandwidth is sold to premium users (Luminati~\cite{luminati}) or used for malicious purposes~\cite{holaDDos}.

VPN Gate~\cite{vpnGate} is a service that aims to achieve blocking resistance to censorship firewalls such as the Great Firewall of China.  Classic VPNs easily fail at this task because their limited and static network footprint can be easily blocked (IP blacklisting). The rationale of VPN Gate is to build a dVPN atop of volunteer machines, and realize a large set of dynamic IP addresses. The authors further inject innocent IP addresses in their public IP lists which makes it harder to perform large IP blacklist. Additionally, they allow their VPN nodes to cooperate in order to quickly identify a list of spies, or computers used by censorship authorities to probe the volunteer dVPN nodes. 

With the recent rise of blockchain, a new form of dVPNs has surfaced. In such, the rationale is to share a user’s upload bandwidth in exchange for some crypto tokens. A popular such example is Mysterium~\cite{mysterium}: an open source dVPN completely built upon a P2P architecture. An immutable smart contract running on Ethereum will be used to make sure that the VPN service is paid adequately. It is currently in alpha test, and incentives will be available soon. In addition, Sentinel~\cite{sentinel} is a larger project of which a dVPN is just one of the use cases. The main idea here is to use the blockchain to store a ledger of data transactions with a \emph{Proof of Traffic}. According to Sentinel's Lite Paper~\cite{sentinelPaper} the ``liability of traffic at the exit node is also upon the host``.

In Section~\ref{sec:requirements}, we describe in detail the requirements a dVPN needs to fulfill. After exploring above the existing real world implementations, in Table~\ref{tab:vpn_summ}, we benchmark the existing dVPNs implementations with respect to the set requirements while also comparing with \tool. Note that this benchmarking was derived from the public information available about existing dVPNs. In the table, we label as \emph{research} the older approaches discussed at the beginning of the section: ELA, N2N, and SocialVPN. 

\begin{table}[t]
  \small
  \centering
  \begin{tabular}{r|ccp{.4cm}p{1.5cm}|c}
    \toprule
    {\bf Requirements} & {\bf Research} & {\bf Hola}   & {\bf VPN Gate} & {\bf Mysterium/ Sentinel} & {\bf \tool}  \\
    \hline
    Open Source         & \checkmark  & \checkmark  & \checkmark  & \checkmark & \checkmark \\
    IP Blacklisting        & \checkmark & X  & \checkmark  & \checkmark & \checkmark \\
    QoS Guarantees      &  X & X  & X  & X  & -\\
    No Logging           &  X & X  & X  & X & \checkmark\\
    Traffic Account   & X & X  & X  & \checkmark & \checkmark\\
    Traffic Blame         & X & X  & X  & X &  \checkmark \\
    \bottomrule
    \hline
  \end{tabular}
  \caption{Comparison of \tool versus the existing approaches with respect to the set requirements.} \vspace{-0.3cm}
  \label{tab:vpn_summ}
\end{table} 

\vspace{0.1in}
\section{Conclusion and Future Work}
\label{sec:conclusion}
Virtual private networks (VPNs) are widely adopted solutions to protect user privacy, circumvent censorship, and access geo-filtered content. This paper has investigated  decentralized VPNs (dVPNs), recent VPN solutions where users are both VPN clients and nodes. Overall,
existing dVPN designs fail to provide strong privacy guarantees. Most notably,
the decentralized nature requires strong guarantees on the traffic a dVPN node
carries without violating a user's privacy, at any time.  To tackle this
problem we designed \emph{\tool}, to the best of our knowledge, the first dVPN
with strong privacy requirements and high performance. \tool is built around a
Distributed Hash Table (DHT) atop of which we designed several privacy
preserving mechanisms. Most notably, a strategy to proof, in zero knowledge,
that a dVPN client is attempting to access a TLS domain. This information is
used by dVPN nodes to only allow through them traffic they are willing to
carry, without violating dVPN users privacy. We integrated \tool with
BitTorrent DHT and ProtonVPN, and benchmarked its performance. Our preliminary
results show the feasibility of the approach but they also highlight a need of
more research to speed up \tool's zero knowledge calculations. We believe the
strong privacy guarantees offered by \tool will foster the development of
future VPN solutions and protocols. 
\newpage

\bibliographystyle{abbrv} 
\balance
\begin{small}

\begin{thebibliography}{10}

\bibitem{hola}
Hola free vpn - unblock any website.
\newblock \url{https://hola.org/}.

\bibitem{ownvpn}
Create your own vpn.
\newblock \url{https://proprivacy.com/guides/create-your-own-vpn-server}, 2019.

\bibitem{thepiratebay}
Anonymous.
\newblock The pirate bay.
\newblock \url{https://www.thepiratebay.org}, 2019.

\bibitem{1386101}
S.~{Aoyagi}, M.~{Takizawa}, M.~{Saito}, H.~{Aida}, and H.~{Tokuda}.
\newblock Ela: a fully distributed vpn system over peer-to-peer network.
\newblock In {\em The 2005 Symposium on Applications and the Internet}, pages
  89--92, Feb 2005.

\bibitem{interval:efficient}
F.~Boudot.
\newblock Efficient proofs that a committed number lies in an interval.
\newblock In B.~Preneel, editor, {\em Advances in Cryptology --- EUROCRYPT
  2000}, pages 431--444, Berlin, Heidelberg, 2000. Springer Berlin Heidelberg.

\bibitem{holaAttack}
M.~Brinkmann.
\newblock Beware: Hola vpn turns your pc into an exit node and sells your
  traffic.
\newblock
  \url{https://www.ghacks.net/2015/05/28/\\beware-hola-vpn-turns-your-pc-into-an-exit-node-and-sells-your-traffic/},
  2015.

\bibitem{camenisch:safeprimes}
J.~Camenisch and M.~Michels.
\newblock Proving in zero-knowledge that a number is the product of two safe
  primes.
\newblock In J.~Stern, editor, {\em Advances in Cryptology --- EUROCRYPT '99},
  pages 107--122, Berlin, Heidelberg, 1999. Springer Berlin Heidelberg.

\bibitem{Camenisch:1997:EGS:646762.706305}
J.~Camenisch and M.~Stadler.
\newblock Efficient group signature schemes for large groups (extended
  abstract).
\newblock In {\em Proceedings of the 17th Annual International Cryptology
  Conference on Advances in Cryptology}, CRYPTO '97, pages 410--424, London,
  UK, UK, 1997. Springer-Verlag.

\bibitem{cholez2009evaluation}
T.~Cholez, I.~Chrisment, and O.~Festor.
\newblock Evaluation of sybil attacks protection schemes in kad.
\newblock In {\em IFIP International Conference on Autonomous Infrastructure,
  Management and Security}, pages 70--82. Springer, 2009.

\bibitem{ciao_url}
{CIAO Team.}
\newblock {Automated free proxies discovery/usage}.
\newblock
  \url{https://chrome.google.com/webstore/detail/automated-free-proxies-di/ojjklffhhhfpeaelghfocilljceokage?hl=en}.

\bibitem{dnscrypt}
D.~Crypt.
\newblock Dnscrypt.
\newblock \url{https://dnscrypt.info/protocol/}, 2008.

\bibitem{Dittus:2018:PCL:3178876.3186094}
M.~Dittus, J.~Wright, and M.~Graham.
\newblock Platform criminalism: The 'last-mile' geography of the darknet market
  supply chain.
\newblock In {\em Proceedings of the 2018 World Wide Web Conference}, WWW '18,
  pages 277--286, Republic and Canton of Geneva, Switzerland, 2018.
  International World Wide Web Conferences Steering Committee.

\bibitem{elgamal}
T.~El~Gamal.
\newblock A public key cryptosystem and a signature scheme based on discrete
  logarithms.
\newblock In {\em Proceedings of CRYPTO 84 on Advances in Cryptology}, pages
  10--18, New York, NY, USA, 1985. Springer-Verlag New York, Inc.

\bibitem{snidraft}
E.~et~al.
\newblock Tls 1.3 sni encrypt.
\newblock \url{https://tools.ietf.org/html/draft-ietf-tls-esni-03}, 2019.

\bibitem{emule}
M.~et~al.
\newblock Emule.
\newblock \url{https://www.emule-project.net/home/perl/general.cgi?l=1}, 2019.

\bibitem{puppeteer}
Google.
\newblock Puppeteer.
\newblock \url{https://github.com/GoogleChrome/puppeteer}, 2019.

\bibitem{sentinelPaper}
S.~S. Group.
\newblock Sentinel - lite paper.
\newblock
  \url{https://github.com/sentinel-official/sentinel/blob/master/README.md},
  2018.

\bibitem{10.1007/3-540-45961-8_11}
L.~C. Guillou and J.-J. Quisquater.
\newblock A practical zero-knowledge protocol fitted to security microprocessor
  minimizing both transmission and memory.
\newblock In {\em Advances in Cryptology --- EUROCRYPT '88}, pages 123--128,
  Berlin, Heidelberg, 1988. Springer Berlin Heidelberg.

\bibitem{harris1999tcp}
B.~Harris and R.~Hunt.
\newblock Tcp/ip security threats and attack methods.
\newblock {\em Computer communications}, 22(10):885--897, 1999.

\bibitem{ikram2016analysis}
M.~Ikram, N.~Vallina-Rodriguez, S.~Seneviratne, M.~A. Kaafar, and V.~Paxson.
\newblock An analysis of the privacy and security risks of android vpn
  permission-enabled apps.
\newblock In {\em Proceedings of the 2016 Internet Measurement Conference},
  pages 349--364. ACM, 2016.

\bibitem{tor}
T.~P. Inc.
\newblock Tor project.
\newblock \url{https://www.torproject.org/}, 2006.

\bibitem{ispact}
M.~Jackson.
\newblock Ipact – controversial new uk isp internet snooping bill becoming
  law.
\newblock
  \url{https://www.ispreview.co.uk/index.php/2016/11/controversial-new-uk-internet-snooping-bill-approved-mps.html},
  2016.

\bibitem{juste2010socialvpn}
P.~S. Juste, D.~Wolinsky, P.~O. Boykin, M.~J. Covington, and R.~J. Figueiredo.
\newblock Socialvpn: Enabling wide-area collaboration with integrated social
  and overlay networks.
\newblock {\em Computer Networks}, 54(12):1926--1938, 2010.

\bibitem{kaashoek2003koorde}
M.~F. Kaashoek and D.~R. Karger.
\newblock Koorde: A simple degree-optimal distributed hash table.
\newblock In {\em International Workshop on Peer-to-Peer Systems}, pages
  98--107. Springer, 2003.

\bibitem{ispAdvertisers}
J.~Kastrenakes.
\newblock Congress just cleared the way for internet providers to sell your web
  browsing history.
\newblock
  \url{https://www.cnbc.com/2017/03/28/congress-clears-way-for-isps-to-sell-browsing-history.html},
  2017.

\bibitem{khan2018empirical}
M.~T. Khan, J.~DeBlasio, G.~M. Voelker, A.~C. Snoeren, C.~Kanich, and
  N.~Vallina-Rodriguez.
\newblock An empirical analysis of the commercial vpn ecosystem.
\newblock In {\em Proc. ACM IMC}, 2018.

\bibitem{luminati}
{Luminati Networks Ltd.}
\newblock Luminati: World's largest proxy service.
\newblock \url{https://luminati.io/}.

\bibitem{holaDDos}
A.~J. Martin.
\newblock Do you use hola vpn? you could be part of a ddos, content theft –
  or worse.
\newblock
  \url{https://www.theregister.co.uk/2015/06/10/hola\_gets\_holes\_poked\_in\_client\_lulzsec/},
  2015.

\bibitem{vpnGate}
D.~Nobori and Y.~Shinjo.
\newblock {VPN} gate: A volunteer-organized public {VPN} relay system with
  blocking resistance for bypassing government censorship firewalls.
\newblock In {\em Proceedings of the 11th {USENIX} Symposium on Networked
  Systems Design and Implementation ({NSDI} 14)}, pages 229--241, Seattle, WA,
  2014. {USENIX}.

\bibitem{openvpn}
OpenVPN.
\newblock Open vpn.
\newblock \url{https://openvpn.net/}, 2019.

\bibitem{Peng:2010:ERP:1906497.1907394}
K.~Peng and F.~Bao.
\newblock An efficient range proof scheme.
\newblock In {\em Proceedings of the 2010 IEEE Second International Conference
  on Social Computing}, SOCIALCOM '10, pages 826--833, Washington, DC, USA,
  2010. IEEE Computer Society.

\bibitem{proton}
ProtonVPN.
\newblock Proton vpn.
\newblock \url{https://protonvpn.com/}, 2019.

\bibitem{tls13}
E.~Rescorla.
\newblock Tls 1.3 rfc.
\newblock \url{https://tools.ietf.org/html/rfc8446}, 2018.

\bibitem{sentinel}
{Sentinel.co}.
\newblock Sentinel: Interoperable network layer for distributed resources.
\newblock \url{https://sentinel.co/}.

\bibitem{adiosHola}
{slipstream/RoL, D. O'Cearbhaill, S. Slootweg, IceMans/RoL, infodox,
  pathfinder/braenaru, APT1337, spoonzy, LeShadow,}.
\newblock Adios, hola! or: Why you should immediately uninstall hola.
\newblock \url{http://adios-hola.org/}, 2015.

\bibitem{steiner2010evaluating}
M.~Steiner, D.~Carra, and E.~W. Biersack.
\newblock Evaluating and improving the content access in kad.
\newblock {\em Peer-to-peer networking and applications}, 3(2):115--128, 2010.

\bibitem{tls12}
E.~R. T.~Dierks.
\newblock Tls 1.2 rfc.
\newblock \url{https://tools.ietf.org/html/rfc5246}, 2008.

\bibitem{mysterium}
{The Mysterium Network}.
\newblock Mysterium network: Decentralised vpn built on blockchain.
\newblock \url{https://mysterium.network/}.

\bibitem{transmission}
TransmissionBT.
\newblock Transmission.
\newblock \url{https://transmissionbt.com/}, 2019.

\bibitem{varvello2011traffic}
M.~Varvello and M.~Steiner.
\newblock Traffic localization for dht-based bittorrent networks.
\newblock In {\em International Conference on Research in Networking}, pages
  40--53. Springer, 2011.

\bibitem{zhao2004tapestry}
B.~Y. Zhao, L.~Huang, J.~Stribling, S.~C. Rhea, A.~D. Joseph, and J.~D.
  Kubiatowicz.
\newblock Tapestry: A resilient global-scale overlay for service deployment.
\newblock {\em IEEE Journal on selected areas in communications}, 22(1):41--53,
  2004.

\end{thebibliography}

\end{small}
\end{document}